\documentclass[aip,apl,twocolumn,superscriptaddress]{revtex4}
\usepackage{graphicx}
\usepackage{morefloats}
\usepackage{color}
\usepackage{amssymb}
\usepackage{float}

\begin{document}

\newcommand{\Py}{Ni$_{80}$Fe$_{20}$} 


\title{Spin torque ferromagnetic resonance with magnetic field modulation}


\author{A.\,M.\,Gon\c{c}alves}
\affiliation{Physics and Astronomy, University of California, Irvine, CA 92697}
\affiliation{Centro Brasileiro de Pesquisas F\'{i}sicas, Rua Dr.\,Xavier Sigaud, 150, Rio de Janeiro 22.290-180, RJ, Brazil}
\author{I.\,Barsukov}
\affiliation{Physics and Astronomy, University of California, Irvine, CA 92697}
\author{Y.-J.\,Chen}
\affiliation{Physics and Astronomy, University of California, Irvine, CA 92697}
\author{L.\,Yang}
\affiliation{Physics and Astronomy, University of California, Irvine, CA 92697}
\author{J.\,A.\,Katine}
\affiliation{HGST Research Center, San Jose, CA 95135}
\author{I.\,N.\,Krivorotov}
\affiliation{Physics and Astronomy, University of California, Irvine, CA 92697}



\begin{abstract}
We demonstrate a technique of broadband spin torque ferromagnetic resonance (ST-FMR) with magnetic field modulation for measurements of spin wave properties in magnetic nanostructures. This technique gives great improvement in sensitivity over the conventional ST-FMR measurements, and application of this technique to nanoscale magnetic tunnel junctions (MTJs) reveals a rich spectrum of standing spin wave eigenmodes. Comparison of the ST-FMR measurements with micromagnetic simulations of the spin wave spectrum allows us to explain the character of low-frequency magnetic excitations  in nanoscale MTJs.\\ \\
The following article appeared in \textit{Appl. Phys. Lett.} \textbf{103}, 172406 (2013) and may be found at http://dx.doi.org/10.1063/1.4826927\\
Copyright (2013) American Institute of Physics. This article may be downloaded for personal use only. Any other use requires prior permission of the author and the American Institute of Physics.
\end{abstract}



\maketitle

Nanoscale magnetic tunnel junctions are promising candidates for non-volatile spin torque memory (STT-RAM) \cite{Slonczewski1989,Huai2004,Fuchs2004,Katine2008,Brataas2012} and current-controlled microwave oscillators  \cite{Kiselev2003,Rippard2004, Deac2008,Zhou2009,Quinsat2010,Demidov2010}. Optimization of MTJ performance for these applications requires quantitative evaluation of the spin torque (ST) vector \cite{Sankey2007}, magnetic damping\cite{Nembach2013}, voltage-controlled magnetic anisotropy \cite{Maruyama2009,Wang2012,Zhu2012}, and the spectrum of magnetic excitations of the MTJ \cite{Helmer2010,Naletov2011}. A common techniques for quantitative measurements of these properties is ST ferromagnetic resonance (ST-FMR) \cite{Tulapurkar2005,Sankey2006}. 

There are three main realizations of ST-FMR: (i) time domain \cite{Wang2011}, (ii) network analyzer \cite{Xue2012}, and (iii) rectification \cite{Tulapurkar2005} methods. The first two methods employ non-zero bias current while the rectification technique allows studies of MTJ magneto-dynamics at zero bias. In the rectification ST-FMR, a microwave current $I_{ac}\cos(2\pi f t)$ flowing across the MTJ applies ST to the magnetization and thereby induces resistance oscillations $R_{ac}\cos(2\pi f t+\psi)$. Mixing of the current and resistance oscillations results in a direct voltage  $V_{mix}=\frac{1}{2}I_{ac}R_{ac}\cos(\psi)$ generated by the sample. Measurements of the ST-FMR spectrum $V_{mix}(f)$ give a series of spectral peaks arising from spin wave resonances of the MTJ \cite{Tulapurkar2005,Sankey2006,McMichael2005}, which can give comprehensive information on voltage-driven magneto-dynamics in the sample. Despite its popularity, the rectification technique suffers from frequency-dependent background signals due to non-linearities and impedance mismatches within the  microwave circuit. For example, measurements of MTJ with collinear free and pinned layer magnetizations (the STT-RAM geometry) are challenging with the rectification ST-FMR because magnetic signals are typically weaker than the backgrounds.    

In this Letter we present a simple technique that circumvents the drawbacks of the conventional ST-FMR and gives reliable data for an arbitrary magnetic state of the MTJ, including the collinear configuration. We solve the problem of the parasitic backgrounds by using magnetic field modulation. The modulation field of a few Oersteds is applied to the sample by augmenting a conventional ST-FMR setup with a copper wire that carries kHz-range sinusoidal current of a few Amperes and is placed directly above the MTJ sample as shown in Fig.\,1(a). The modulation field from the wire is collinear with the dc applied magnetic field at the sample location. A continuous microwave current is applied to the sample via a bias tee, and a rectified voltage generated by the sample is measured by a lock-in amplifier at the field modulation frequency. For reference, we also make conventional ST-FMR measurements in which the sinusoidal magnetic field modulation is replaced by a square wave amplitude modulation of the microwave current.
\begin{figure*}
\includegraphics[width=2.0\columnwidth]{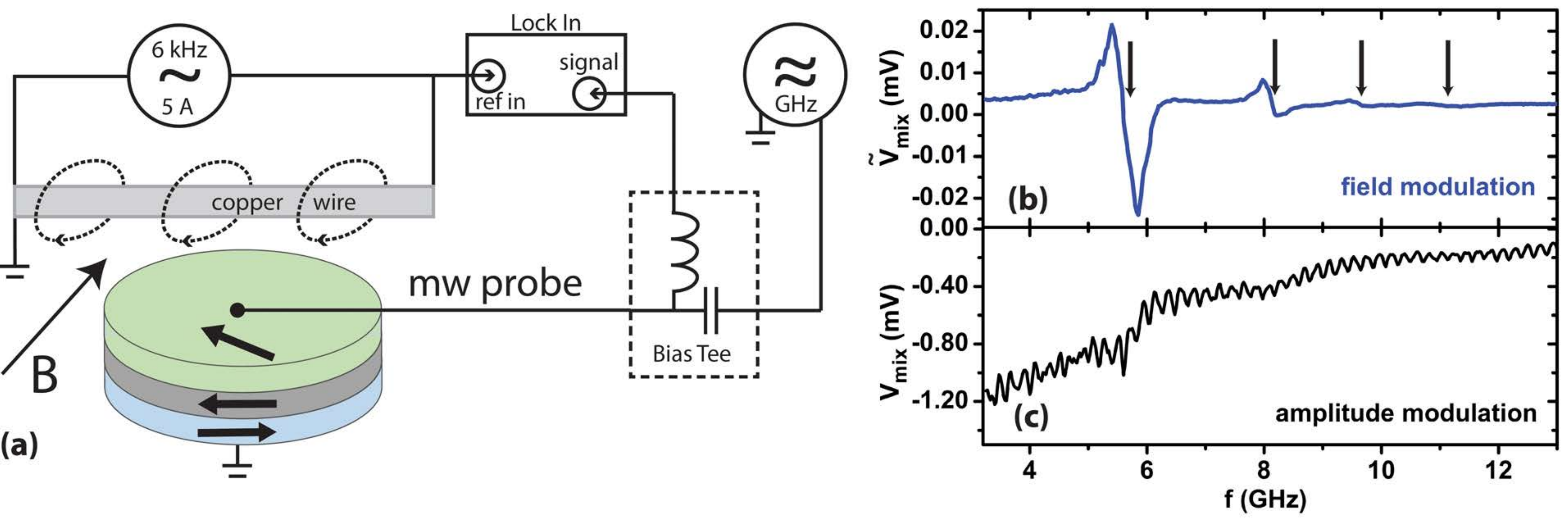}
\caption{\label{exp}(Color online) (a)~Sketch of the field-modulation ST-FMR setup. Comparison of ST-FMR spectra measured by (b) field and (c) amplitude modulation techniques for external magnetic field of 30\,mT applied along the easy axis of a 170$\times$90\,nm$^2$ elliptical MTJ nanopillar. Field modulation efficiently removes large frequency-dependent background of non-magnetic origin and reveals four spin wave eigenmodes of the MTJ.}
\end{figure*}  

We make ST-FMR measurements of magnetization dynamics in a 170$\times$90\,nm$^2$ elliptical MTJ nanopillar with in-plane magnetizations of the pinned and free layers.  The nanopillar is patterned by ion milling from a Ta(5)/SAF/MgO(1.024)/FL/Ta(5) multilayer (thicknesses in nm) with resistance-area product of 14 $\Omega\cdot\mu$m$^2$ deposited by magnetron sputtering in a Singulus TIMARIS system. Here SAF =  PtMn(15)/ Co$_{70}$Fe$_{30}$(2.5)/ Ru(0.85)/ Co$_{40}$Fe$_{40}$B$_{20}$(2.4) is the synthetic antiferromagnet pinned layer and FL = Co$_{60}$Fe$_{20}$B$_{20}$(1.8)  is the free layer.  Prior to patterning, the multilayers are annealed for 2 hours at 300\,$^\circ$C in a 1 Tesla in-plane magnetic field that sets the pinned layer exchange bias direction parallel to the long axis of the nanopillar. For evaluation of the field-modulated ST-FMR method, we employ the collinear configuration, in which an external magnetic field of 30\,mT is applied along the easy axis of the nanopillar. This configuration is particularly unfavorable for ST-FMR measurements because (i) ST proportional to the sine of the angle between the pinned and the free layer magnetizations is small and (ii) resistance oscillations at the frequency of the microwave drive are small. Not surprisingly, the ST-FMR spectrum for this configuration measured with the amplitude modulation technique (Fig.\,1(c)) is dominated by a frequency-dependent non-magnetic background. In contrast, field-modulated ST-FMR spectrum shown in Fig.\,1(b) is nearly background-free and reveals four resonances arising from excitation of spin wave eigenmodes of the MTJ. 

\begin{figure*}
\includegraphics[width=2.0\columnwidth]{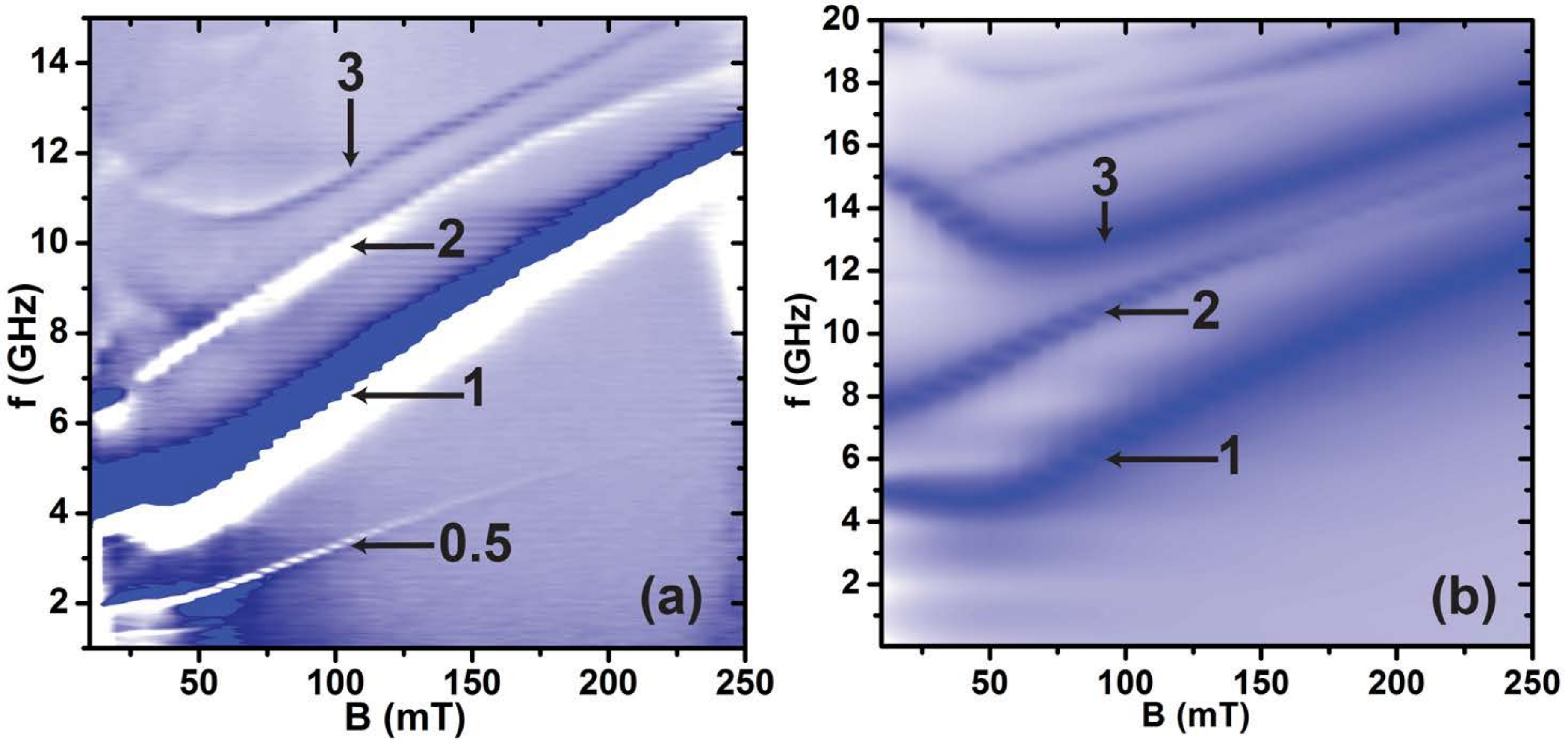}
\caption{\label{3d}(Color online) (a)~Field-modulated ST-FMR spectra measured for an 85$\times$50\,nm$^2$ elliptical MTJ nanopillar as a function of hard-axis magnetic field. (b)~MTJ spin wave spectra versus hard-axis magnetic field derived from micromagnetic simulations. The color plot shows FFT amplitude of the MTJ out-of-plane magnetic moment excited by a sinc-shaped current pulse.}
\end{figure*}

We next employ the field-modulated ST-FMR for studies of MTJ spin wave eigenmodes as a function of external magnetic field applied in the plane of the sample parallel to the hard axis of the MTJ nanopillar. Fig.\,\ref{3d}(a) gives a color plot summary of ST-FMR spectra measured for an 85$\times$50\,nm$^2$ elliptical MTJ nanopillar as a function of hard-axis magnetic field. As the resonance signals are predominantly antisymmetric (similar to Fig.\,\ref{exp}(b)), the resonance frequencies are found in the vicinity of zero-crossing of the rectified voltage. Three prominent spin wave resonances labeled as 1, 2, and 3 are seen in the entire magnetic field range. The first mode exhibits the highest intensity, lowest frequency, and a field dispersion with a frequency minimum, which is characteristic of the hard-axis quasi-uniform mode. The nature of the higher frequency spin wave modes  is not immediately clear. The frequency of the second mode monotonically increases with the field, while the third mode shows a frequency minimum at a field higher than that of the first mode. We also observe a resonance labeled 0.5 in Fig.\,\ref{3d}(a) for which frequency and linewidth are exactly half of those of the first mode. Therefore, the 0.5 resonance is not a real spin wave mode but an artifact arising from frequency doubling of the current by nonlinearities of the  circuit. 

Next, we derive an expression describing the spectral line shape measured by the field-modulated ST-FMR.  The line shape $V_{mix}(f)$ without field modulation is a sum of symmetric $S(f)$ and antisymmetric $A(f)$ Lorentzians \cite{Wang2009} $V_{mix}(f) = V_s S(f)+ V_a A(f)$, where $S(f)=\frac{1}{1+(f-f_r)^2/\sigma_r^2}$, $A(f)=\frac{ (f-f_r)/\sigma_r}{1+(f-f_r)^2/\sigma_r^2}$, $f_r$ is the resonance frequency and $\sigma_r$ is the linewidth. The amplitudes of the Lorentzians, $V_s$ and $V_a$, are functions of the ST vector, magnetic parameters of the system \cite{Wang2009}, and voltage controlled magnetic anisotropy \cite{Zhu2012}. When the modulation field is small compared to the resonance linewidth in the field domain, the RMS voltage signal $\tilde{V}_{mix}(f)$ measured by the lock-in amplifier is proportional to the first derivative of the rectified voltage $V_{mix}(f)$ with respect to the modulated variable--the external magnetic field $B$:
\begin{widetext}
\begin{eqnarray}
\label{Eq2}
\tilde{V}_{mix}(f) = B_m\frac{\mathrm{d}V_{mix}(f)}{\mathrm{d}B} = B_m\left[\frac{\text{d} V_s}{\text{d} B}S(f)+\frac{\text{d} V_a}{\text{d} B}A(f)+ \frac{1}{\sigma_r}\frac{\mathrm{d}\sigma_r}{\mathrm{d}B}  \left( 2\;V_s\;A^2(f) + V_a \left[ 2\;A^3(f)/S(f)-A(f) \right] \right)\right.\\ \nonumber
\left. +\frac{1}{\sigma_r}\frac{\text{d} f_r}{ \text{d} B}\left(2\,V_s\,S(f)\,A(f)+V_a \left[ A^2(f)-S^2(f) \right]\right) \right],
\end{eqnarray}
\end{widetext}
where $B_m$ is the RMS amplitude of the modulation field, and the last term proportional to $\mathrm{d}f_r/\mathrm{d}B$ is usually dominant. If $V_s$ and $V_a$ are weak functions of magnetic field then the symmetric part of $\tilde{V}_{mix}(f)$ is proportional to $V_a$ and the antisymmetric part is proportional to $V_s$. For our samples, the resonances are nearly antisymmetric and thus $V_s \gg V_a$, which means that the field-like component of ST is much smaller than the in-plane component \cite{Sankey2007,Wang2009}.

To understand the nature of the observed spin wave excitations, we perform OOMMF \cite{OOMMF,McMichael2005} micromagnetic simulations of magnetization dynamics. To fully account for all magnetic interactions in the MTJ, we employ a three dimensional model with three ferromagnetic layers: free, SAF top and SAF bottom. We use material parameters obtained from independent measurements and/or their accepted literature values \cite{MatParam}. In the simulations, spin wave dynamics is excited by a combined pulse of ST and Oersted field, both resulting from a sinc-shaped $J_c \frac{ \sin{ (2 \pi f_{c} t')} }{ 2 \pi f_{c} t' }$ spatially uniform current pulse \cite{Venkat2013} with the amplitude $J_c$ = 9$\times10^6$\,A/cm$^2$, the cut-off frequency $f_{c}=20$\,GHz and the time variable $t'=t-500$\,ps. The spatial profile of the Oersted field is assumed to be that of a long wire with elliptical cross section. The direction of the ST vector acting on the free layer is determined by the magnetization orientation of the SAF top layer.
\begin{figure*}
\includegraphics[width=2.0\columnwidth]{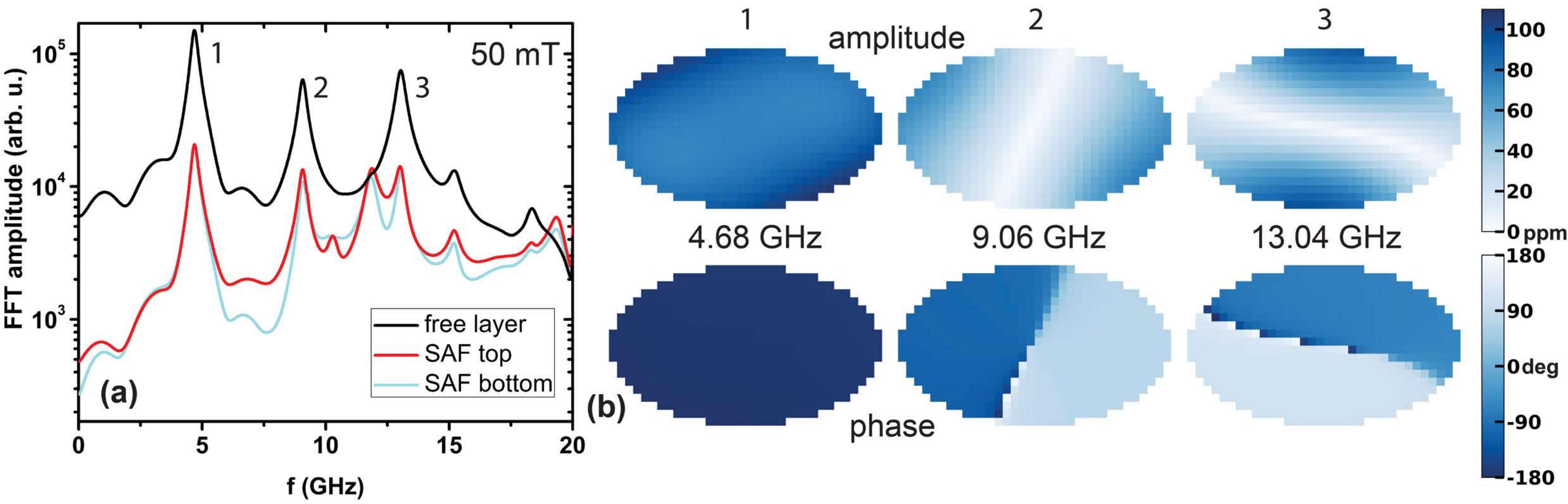}
\caption{\label{sim}(Color online) Micromagnetic simulations of spin wave eigenmodes of 85$\times$50 nm$^2$ MTJ nanopillar at 50\,mT hard axis magnetic field: (a)~Precession amplitude of the free layer, SAF top and bottom layers as a function of frequency. (b)~Color-coded spatial maps of precession amplitude and phase of the three dominant spin wave modes in the MTJ free layer.}
\end{figure*} 
The spectrum of spin wave eigenmodes is obtained via the Fast-Fourier-transform (FFT) of the time dependent out-of-plane component of the MTJ net magnetic moment. The simulated  FFT spectra are shown in Fig.\,\ref{3d}(b) as a function of external magnetic field. Several spin wave modes are clearly seen in Fig.\,\ref{3d}(b), of which three modes stand out due to their large amplitude. The field dispersion of these modes is in a good agreement with the experimental data shown in Fig.\,\ref{3d}(a). The first and third modes have a frequency minimum at approximately 50\,mT and 75\,mT -- values similar to those observed in the experiment. The absence of the 0.5 resonance in the simulation confirms that this feature is an artifact.

To identify the character of the observed spin wave modes, we perform layer- and spatially resolved analysis of the mode amplitude. The frequency dependence of the FFT precession amplitude at 50\,mT field, averaged over all micromagnetic cells of each magnetic layer is shown in Fig.\,\ref{sim}(a). While static magnetic moments of all three layers  are similar to each other, the amplitude of the free layer precession is significantly higher than that of the SAF layers. Thus we classify the observed spin wave excitations as the free layer modes. Analysis of spatial distribution of the amplitude and phase of the excitations in the free layer, shown in Fig.\,\ref{sim}(b), reveals the nature of these spin wave modes. The amplitude and the phase of the first mode are homogeneous across the entire free layer, which means that this mode is the free layer quasi-uniform mode. The second mode has maximum excitation amplitude at the edges of the free layer with a node along the short axis. The phase of the oscillations undergoes a 180$^\circ$ shift across the node indicating that the second mode is antisymmetric. The spatial profile of the third mode is qualitatively similar to that of the second mode, but the node is along the long axis of the ellipse. Deviations of the nodes from the symmetry axes of the ellipse are due to a stray field from the SAF layer. Given the imperfections of the real MTJ structure and uncertainties in the MTJ material parameters, the energies of the spin wave eigenmodes found in simulations are in good agreement with the experimental results. 
 
In conclusion, we developed a method of field-modulated ST-FMR for characterization of magnetization dynamics in  nanostructures and compared it to the conventional amplitude-modulated ST-FMR. The field-modulated ST-FMR technique suppresses large non-magnetic background signals and thereby reveals a rich spectrum of spin wave eigenmodes. Using the field-modulated ST-FMR, we measured the spectrum of spin waves in nanoscale elliptical MTJs as a function of hard-axis magnetic field. Comparison of the ST-FMR data with micromagnetic simulations shows that the observed  modes are low-energy  eigenmodes of the free layer. 

We thank J.\,Langer for magnetic multilayer deposition and R.\,Arias for helpful discussions. This work was supported by SRC and Intel through grant No. 2011-IN-2152  as well as by NSF through Grants No. DMR-1210850, No. DMR-0748810, and No. ECCS-1002358.  A.\,M.\,G. thanks CAPES Foundation, Ministry of Education of Brazil for funding through process number 4977-11-4.

\end{document}